# Automatic Plane Adjustment of Orthopedic Intraoperative Flat Panel Detector CT-Volumes[*]

Celia Martín Vicario[1,2], Florian Kordon[1,2,3][0000-0003-1240-5809], Felix Denzinger[1,2], Markus Weiten[2], Sarina Thomas[4], Lisa Kausch[4], Jochen Franke[5], Holger Keil[5], Andreas Maier[1,3][0000-0002-9550-5284] and Holger Kunze[2][0000-0002-7021-2370]

[1] Pattern Recognition Lab, Friedrich-Alexander-Universität Erlangen-Nürnberg (FAU), Erlangen, Germany
[2] Siemens Healthcare GmbH, Forchheim, Germany
[3] Erlangen Graduate School in Advanced Optical Technologies (SAOT), Friedrich-Alexander-Universität Erlangen-Nürnberg (FAU), Erlangen, Germany
[4] Division of Medical Image Computing, German Cancer Research Center, Heidelberg, Germany
[5] Department for Trauma and Orthopaedic Surgery, BG Trauma Center Ludwigshafen, Ludwigshafen, Germany
`Holger.HK.Kunze@siemens-healthineers.com`

**Abstract.** Flat panel computed tomography is used intraoperatively to assess the result of surgery. Due to workflow issues, the acquisition typically cannot be carried out in such a way that the axis aligned multiplanar reconstructions (MPR) of the volume match the anatomically aligned MPRs. This needs to be performed manually, adding additional effort during viewing the datasets. A PoseNet convolutional neural network (CNN) is trained such that parameters of anatomically aligned MPR planes are regressed. Different mathematical approaches to describe plane rotation are compared, as well as a cost function is optimized to incorporate orientation constraints. The CNN is evaluated on two anatomical regions. For one of these regions, one plane is not orthogonal to the other two planes. The plane's normal can be estimated with a median accuracy of 5°, the in-plane rotation with an accuracy of 6°, and the position with an accuracy of 6 mm. Compared to state-of-the-art algorithms the labeling effort for this method is much lower as no segmentation is required. The computation time during inference is less than 0.05 s.

**Keywords:** Orthopedics · Flat Planel CT · Multiplanar Reconstruction

## 1 Introduction

Intraoperative 3D acquisition is an important tool in trauma surgery for assessing the fracture reduction and implant position during a surgery and the result before releasing the patient out of the operating room [1]. While X-ray images help to assess the result of surgeries in standard cases, in complex anatomical regions like the calcaneus or ankle, 3D imaging provides a mean to resolve ambiguities. Recent studies have shown that, depending on the type of surgery, intraoperative correction rates are up to 40% when such systems are available [2]. If such a tool is not available, postoperative computed tomography is recommended. If a result is observed that should be revised (e.g. an intraarticular screw misplacement) a second surgery is necessary. Therefore, intraoperative 3D scans help to improve the outcome of the surgery and to avoid revision surgeries.

---

[*] The authors gratefully acknowledge funding of the Erlangen Graduate School in Advanced Optical Technologies (SAOT) by the Bavarian State Ministry for Science and Art.

To be able to understand the patient's anatomy, the volume should be oriented in a standardized way, as is customary in the volumes provided by the radiology department. For those acquisitions, the patient is typically carefully positioned, so that for example the axial slices of the computed tomography (CT) are aligned to the bone of interest. In the axis aligned multiplanar reconstructions (MPR), steps and gaps especially in intra-articular spaces can be analyzed without any further reformation. Also, with these carefully aligned slices, malpositioned screws can be diagnosed easily.

However, intraoperative 3D acquisitions are performed with the patient lying on the table in the position that the surgery requires. Moreover, in some cases the imaging system cannot be aligned to the patient due to mechanical restrictions and the setup of the operating theater. Consequently, the scan often results in volumes which are not aligned to the patient anatomy. To obtain the correct presentation of the volume, it is essential to correct the rotation of the volume [2]. For the calcaneus region, the surgeon needs an average of 46 to 55 seconds for manual adjustment of the standard planes depending on his experience [3]. Additional time is spent as he gets unsterile and needs at least to change gloves.

For 3D acquisitions often mobile and fixed mounted C-arm systems are used, as they combine the functionality of 2D X-ray imaging and 3D acquisition without adding another device to the operating theater, contrary to intraoperative CT systems. However, their 3D field of view is limited so that they typically cover the region of interest but not more. Often in such volumes only a restricted number of landmarks is visible, therefore landmark based approaches generally fail.

Since the volumes are acquired after a trauma surgery, typically a larger number of metal implants and screws are inserted into the patient's body generating severe metal artifacts. An automation of anatomically correct alignment of axial, coronal, and sagittal MPRs given the above-mentioned restriction is desired.

Literature review shows that automatic alignment is not a new topic. [4] already implemented an automatic rotation of axial head CT slices. [5, 6] covered the derivation of the brain midline using tools of pattern recognition. In [2] SURF features are extracted from the dataset and registered to an atlas with annotated MPRs. The quality is dependent on the choice of atlas and the feature extraction. Also, the registration needs to be carefully designed to support strongly truncated volumes [7] and typically the capture range of rotation is limited. An alternative approach to the present problem was followed by [8], in which shape models with attached labels for MPRs were registered to the volume. Although this solution solves the problem, in order to train the shape models for each body region the anatomical structures need to be segmented in the training datasets. Thus, the use of that algorithm requires a huge amount of manual work. For this approach, a time of 23s was reported for the shape model segmentation and subsequent plane regression.

A first artificial intelligence approach for ultrasound images was presented in [9], in which the plane regression task was solved by using random forest trees. Recently, a convolutional neural network (CNN) was proposed, which extracts the standard slices from a stack of ultrasound images [10]. In that paper the rotation was described using different representations as quaternion or Euler angles where quaternions were showed to be superior. In [11] a rotation matrix was used of which only the first five or six values are calculated (5D or 6D method). This approach applies the definition of a proper rotation matrix, that states that the columns of a rotation matrix are orthonormal vectors and form a right-handed coordinate system. So, knowing the first vector and two coordinates of the second vector suffices to calculate the remaining four entries.

An inter-rater study was performed to evaluate the accuracy of two raters in the manual adjustment of the standard planes for the proximal femur regions in [12]. The found error estimations in that region are 6.3° for the normal and in-plane rotation and 9.3mm for the translation. For this region, the planes are similarly well defined as for the upper ankle region and better defined compared to the calcaneus region.

In this paper we want to present a 3D CNN, with which the standard plane parameters are regressed directly from the volume. Since a plane regression network is used, no segmentation is needed, and the training of the proposed algorithm requires only the description of the planes. We compare three different approaches to describe the rotation and evaluate the optimal strategy for a regression of multiple MPRs.

To the best of our knowledge we present the first 3D CNN to directly estimate standard plane's parameters. Doing so we introduce the 6D method in medical imaging and compare it to well-established methods like Euler angles and quaternions.

In Section 2 we explain the methods of our approach, we introduce the employed mathematical description of planes, describe the normalization of the coordinate system, neural network, and the cost function. The implementation and the data we used for training and testing is described in Section 3 as well as the study design we followed. Thereafter we present the results of our experiments. Finally, we discuss the results in Section 5.

## 2 Methods

### 2.1 Plane Description

Mathematically a plane can be described by the point $A$ and the vectors $e_u$ and $e_v$ showing right and upwards with increasing screen coordinate values. This description has the advantage that the image rotation is incorporated. The plane normal $e_w$ is the cross product of $e_u$ and $e_v$. The point $A$ is the center of the plane.

The matrix $R = [e_u, e_v, e_w]$ can be interpreted as the rotation matrix from the volume to the plane. This rotation can be represented by rotation matrices (6D method), defined by Euler angles as well as by quaternions. As the coordinates of quaternions and matrices are in the range of $[-1, 1]$ these values are used directly. For the Euler angles however, the sine and cosine values of the angles are used. By this, we solve the problem of periodicity of the values and compress the values the range to the interval $[-1, 1]$. The angle value can be retrieved by the atan2 method.

### 2.2 Neural Network and Augmentation

To regress the parameters, we use an adapted version of the CNN proposed in [13] (Fig. 1). In contrast to the original work, the dropout layer was removed.

For each regressed value, the last fully connected layer has one output node. The number per MPR plane depends on the selected model for the rotation and varies between seven for quaternions and nine for sine and cosine values of the Euler angles and 6D representation. Thereby, 3 output nodes represent the translation whereas the remaining nodes describe the rotation.

During training, online augmentation of the volumes is employed. Since a neural network which combines rotation and translation parameters shall be trained using a combined loss function, we normalize the translation such that the origin of the volume is in the center of the volume, and that the volume edge length has a normalized length of 1. We apply a random rotation within the interval $[-45, 45]°$, random spatial scale of the volume by a factor in range $[0.95, 1.05]$, translation by $[-12, 12]$mm, a center crop, and sub-sampling. Additionally, mirroring in x-direction is added with a probability of 0.5 that allows to simulate left-right handedness of the volume.

For speedup and for reducing the number of interpolations, the augmentation operations are applied in a single step using their homogeneous matrices to create a composite matrix and therefore, the spatially augmented volume is interpolated just once during the sub-sampling.

To be robust to imperfect intensity calibration, the HU values added by 1000 HU are multiplied by a factor uniformly sampled of the range $[0.95, 1.05]$. Finally, a windowing function $w(x) = 1/(1 + e^{g(0.5-x)})$ is applied after clipping the volume intensity values to the range of $[-490, 1040]$ HU and rescaling it to $[0, 1]$. The gain parameter $g$ is set dependent on the min/max values. This function helps to compress the values to the range $[0,1]$. In contrast to min-max normalization, it reduces the signal variance of metal and air which typically contains little to no information about the plane's parameters. To speed up read-in, we down-sampled the original volumes to volumes of $128^3$ voxels with length 160.25 mm. The cost function is chosen to be

$$L = \alpha L_{rotation} + \beta L_{translation} + \gamma L_{orthogonality} \qquad (1)$$

with $L_{translation}$ being the Euclidian distance of the normalized value of $\boldsymbol{A}$, $L_{rotation}$ the Euclidian distance of the parameters describing the rotation, and $L_{orthogonality}$ the average of the cross products of the MPR normals.

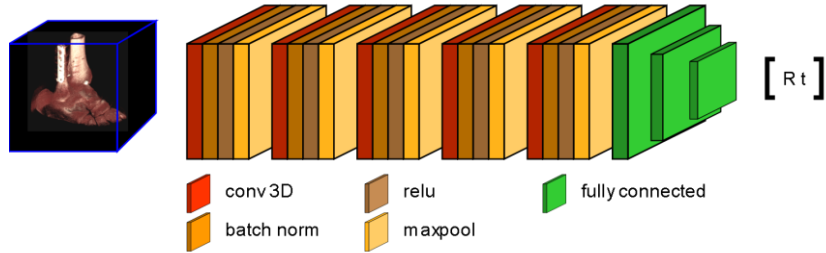

**Fig. 1.** Architecture of the regression CNN. 5 convolutional blocks are followed by 3 fully connected layers.

## 3 Experiments

### 3.1 Datasets

For the evaluation of the approach we use a dataset consisting of 160 volumes of the calcaneus region and 220 volumes of the ankle region. They were partly acquired intraoperatively to assess the result of a surgery and partly from cadavers, which were prepared for surgical training. The cadaver volumes were typically acquired twice: once without modifications and once with surgical instruments laid above, in order to produce metal artifacts without modifying the cadavers. The exact distribution of the datasets is listed in Table 1.

For each body region 5 data splits were created, taking care that volumes of the same patient belong to the same subset and that the distribution of dataset origin is approximately the same as in the total dataset. For all volumes, standard planes were defined according to [13].

For the ankle volumes axial, coronal, and sagittal MPRs, for the calcaneus datasets axial, sagittal, and semi-coronal planes were annotated by a medical engineer after training (Fig. 2). The labelling was performed using a syngo XWorkplace VD20 which was modified to store the plane description. Axial, sagittal, and coronal MPRs were adjusted with coupled MPRs. The semi-coronal plane was adjusted thereafter with decoupled planes.

**Table 1.** Origin and distribution of the used datasets.

|  | Cadaver | | | Clinical | Total |
|---|---|---|---|---|---|
|  | Metal implants | Metal outside | No metal | Metal implants |  |
| Calcaneus | 9 | 63 | 62 | 26 | 160 |
| Ankle | 36 | 61 | 56 | 67 | 220 |

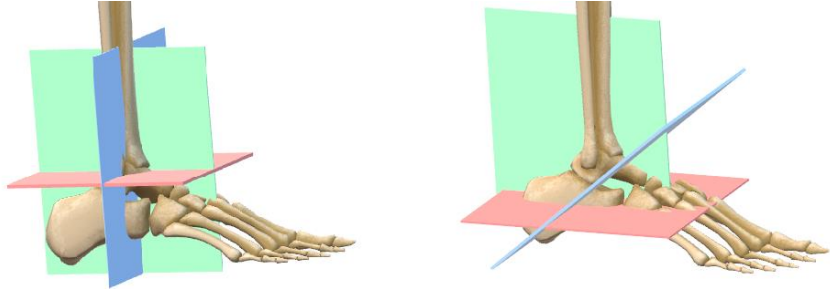

**Fig. 2.** For the dataset the displayed planes were labelled. Left: ankle, right: calcaneus. Red: axial, Green: sagittal, Blue: (semi-)coronal.

### 3.2 Ablation study

We performed an ablation study to find the best configuration of the algorithm. For all comparisons, a 5-fold cross-validation is carried out. As a performance metric we use

$$P = 0.2d + 0.6\varepsilon_n + 0.2\varepsilon_i \quad (2)$$

with $d$ being the median error of the translation in direction of planes normal, $\varepsilon_n$ being the error of the normal vector $e_w$ of the plane, and $\varepsilon_i$ marking the in-plane rotation error calculated as mean difference angle of $e_u$ and $e_v$. The weights were chosen in accordance with our medical advisors and they reflect that the normal orientation is the most important metric, while in-plane rotation and correct plane translation are of subordinate importance, since the rotation does not influence the displayed information and the user scrolls through the volume during the review, actively changing the translation.

First, the influence of the description of the rotation was tested. For both body regions and voxel sizes of 1.2 mm, 2.2 mm, and 2.5 mm with respective volume sizes of $128^3$, $72^3$, $64^3$ voxels, the performance measure for the three different representations was evaluated.

Thereafter, the performance of regressing three planes using one network compared to regressing the planes with separate networks was evaluated. For that the values of $\alpha$, $\beta$, and $\gamma$ were systematically varied, such that $\alpha$ and $\beta$ were in the interval [0.1, 0.9] and $\gamma$ was selected such that $\alpha + \beta + \gamma = 1$. Upon training separate networks for each plane, $\beta$ was selected so that $\alpha + \beta = 1$, not enforcing any orthogonality.

For the cost-function weights optimized in this way, the comparison of using a single network and three separate networks was evaluated.

### 3.3 Implementation

The models are implemented in PyTorch (v.1.2) and trained on Windows 10 systems with 32GB RAM and 24GB NVIDIA TITAN RTX. The weights are initialized by the He et al. method [14]. The network is trained by Stochastic Gradient Descent (SGD) optimizer with momentum. The total number of epochs was set to 400. For selection of learning rate, learning rate decay, step size, momentum, and batch size, a hyper-parameter optimization using random sampling of the search space was performed using one fold and independently for the different rotation descriptions and volume input sizes. The optimization was performed for one of five folds. This method results in an offset of typically 0.1 and maximum 0.4 score points.

## 4     Results

Table 2 shows the results of the evaluation of the error depending on the angle representation. For both body regions, the 6D representation produces the best results for the rotation estimation. Using this method, the error of the normal and the in-plane rotation is minimal. Therefore, the next experiments were carried out using the 6D method.

As can be seen in Table 3, varying the resolution has minor impact on the accuracy. While the training time increases from 20 s over 30 s to 60 s per epoch, the score does not change significantly. Utilizing the small advantage of sampling with 72 voxels, with minor impact on the training time, sampling with 72 voxels was used for the next experiments.

Using a combined network for predicting the parameters of all planes improves the outcome of the network slightly (Table 4). A small further improvement can be reached by adding a constraint on the orthogonality of the planes.

All the experiments show that the plane regression for the ankle region with orthogonal planes works better than for the calcaneus region. Typically, the score is 2 to 3 value points better, with all the error measures contributing in the same way to the improvement. For the comparison of the models, also the error of each plane is listed in Table 5, showing that the orientation of the axial MPR can be estimated best compared to the other MPR orientations. The inference time was below 0.05s.

## 5     Discussion and Conclusion

We have presented an algorithm which allows to automatically regress the standard MPR plane parameters in neglectable time. The proposed algorithm is capable to deal with metal artifacts and strong truncation. Both kind of disturbances are common within intraoperatively acquired volumes.

In contrast to state-of-the-art algorithms, no additional volume segmentation for training or evaluation purposes is needed, consequently reducing the amount of labeling effort needed.

For describing the rotation, using the 6D method is most reliable followed by Euler angles and quaternions. In contrast to [10] and [11], the Euler angles were not regressed directly but their cosine and sine values, yielding a better result compared to quaternions.

We have shown that the algorithm is capable to regress both orthogonal and non-orthogonal planes, with a higher accuracy for the ankle region. A reason for this difference could be that the standard planes for ankle anatomy are better defined and therefore have less variability across samples. Especially the definition of the semi-coronal MPR of the calcaneus allows some variance.

The orthogonality term for the planes normal has a minor benefit for the result. This can be explained with the fact, that the labelled values already follow this constraint, so that it provides only little further information.

We also observe that the combined representation of the plane translational and rotational plane parameters is beneficial for the accuracy of the trained network. Depending on the required accuracy, also a stronger down-sampling allows for clinically sufficient results.

We see a limitation of the evaluation in the missing comparison with the results of [8]. However, due to its need of costly segmentations this was out of scope for this project.

**Disclaimer** The methods and information presented here are based on research and are not commercially available.

**Table 2.** Mean results of plane regression evaluation using different rotation description parameters. Number of voxels is set to $72^3$. The results reflect the mean of the single planes.

|  | Calcaneus | | | | Ankle | | | |
|---|---|---|---|---|---|---|---|---|
|  | d in mm | $\varepsilon_n$ in ° | $\varepsilon_i$ in ° | Score | d in mm | $\varepsilon_n$ in ° | $\varepsilon_i$ in ° | Score |
| **Euler** | 14.39 ±1.64 | 8.93 ±1.60 | 11.05 ±2.10 | 10.45 ±1.18 | 7.78 ±0.36 | 6.99 ±0.81 | 8.37 ±1.23 | 7.42 ±0.77 |
| **Quat.** | 9.93 ±2.25 | 9.96 ±1.75 | 9.54 ±1.59 | 9.87 ±1.66 | 5.00 ±0.09 | 8.16 ±0.79 | 8.31 ±0.71 | 7.56 ±0.63 |
| **6D** | 9.94 ±1.92 | 8.77 ±0.60 | 8.34 ±0.44 | 8.92 ±0.51 | 5.43 ±0.25 | 7.11 ±0.48 | 6.58 ±0.30 | 6.67 ±0.35 |

**Table 3.** Mean results of plane regression evaluation using different numbers of voxels. The rotation is described using the 6D method. The results reflect the mean of the single planes.

|  | Calcaneus | | | | Ankle | | | |
|---|---|---|---|---|---|---|---|---|
|  | d in mm | $\varepsilon_n$ in ° | $\varepsilon_i$ in ° | Score | d in mm | $\varepsilon_n$ in ° | $\varepsilon_i$ in ° | Score |
| $64^3$ | 9.74 ±0.88 | 8.98 ±2.00 | 8.33 ±1.08 | 9.01 ±1.46 | 6.18 ±0.75 | 7.12 ±0.78 | 6.49 ±0.69 | 6.80 ±0.63 |
| $72^3$ | 9.94 ±1.92 | 8.77 ±0.60 | 8.34 ±0.44 | 8.92 ±0.51 | 5.43 ±0.25 | 7.11 ±0.48 | 6.58 ±0.30 | 6.67 ±0.35 |
| $128^3$ | 10.48 ±2.70 | 8.48 ±1.21 | 8.49 ±1.21 | 8.88 ±1.49 | 4.86 ±0.29 | 7.75 ±1.00 | 7.04 ±0.75 | 7.03 ±0.70 |

**Table 4.** Mean results of plane regression evaluation. Three: for each plane an independently optimized model is used (Table 6), Comb.: One model for regressing the parameters for all planes is used with $\alpha = \beta = 0.5$, Opt. Comb.: One model for regressing the parameters for all planes is used with optimized values $\alpha$, $\beta$, and $\gamma$. The rotation is described using the 6D method. Number of voxels is set to $72^3$. The results reflect the mean of the single planes.

|  | Calcaneus | | | | Ankle | | | |
|---|---|---|---|---|---|---|---|---|
|  | d in mm | $\varepsilon_n$ in ° | $\varepsilon_i$ in ° | Score | d in mm | $\varepsilon_n$ in ° | $\varepsilon_i$ in ° | Score |
| **Three** | 9.46 ±1.24 | 9.26 ±0.84 | 8.94 ±0.82 | 9.24 ±0.70 | 6.52 ±0.33 | 7.55 ±0.28 | 6.77 ±0.38 | 7.19 ±0.09 |
| **Comb.** | 9.94 ±1.92 | 8.77 ±0.60 | 8.34 ±0.44 | 8.92 ±0.51 | 5.43 ±0.25 | 7.11 ±0.48 | 6.58 ±0.30 | 6.67 ±0.35 |
| **Opt. Comb.** | 10.38 ±1.89 | 8.14 ±1.21 | 7.91 ±0.79 | 8.54 ±0.83 | 5.20 ±0.21 | 6.93 ±0.58 | 6.86 ±0.36 | 6.57 ±0.44 |

**Table 5.** Regression results for each plane using the combined model with the optimized cost function. The rotation is described using the 6D method. Number of voxels is set to $72^3$.

|  | Calcaneus | | | | Ankle | | | |
|---|---|---|---|---|---|---|---|---|
|  | d in mm | $\varepsilon_n$ in ° | $\varepsilon_i$ in ° | Score | d in mm | $\varepsilon_n$ in ° | $\varepsilon_i$ in ° | Score |
| **Axial** | 10.35 ±3.62 | 7.38 ±1.26 | 7.69 ±0.77 | 8.04 ±1.64 | 6.61 ±0.40 | 5.20 ±0.97 | 7.76 ±0.67 | 5.89 ±0.80 |
| **Semic. /coronal** | 13.11 ±1.21 | 8.71 ±1.12 | 7.49 ±0.89 | 9.35 ±1.09 | 4.56 ±0.54 | 7.57 ±0.77 | 6.05 ±0.37 | 6.67 ±0.35 |
| **Sagittal** | 7.77 ±0.81 | 8.65 ±1.68 | 8.34 ±1.07 | 8.41 ±1.39 | 4.73 ±0.25 | 9.18 ±1.22 | 6.89 ±0.40 | 7.83 ±0.86 |

**Table 6.** $\alpha$, $\beta$, and $\gamma$ values for the different models (Table 4).

|  |  | Calcaneus | | | Ankle | | |
|---|---|---|---|---|---|---|---|
|  |  | $\alpha$ | $\beta$ | $\gamma$ | $\alpha$ | $\beta$ | $\gamma$ |
| **Three** | **Axial** | 0.2 | 0.8 | 0.0 | 0.6 | 0.4 | 0.0 |
|  | **Coronal** | 0.2 | 0.8 | 0.0 | 0.2 | 0.8 | 0.0 |
|  | **Sagittal** | 0.6 | 0.4 | 0.0 | 0.8 | 0.2 | 0.0 |
| **Comb.** |  | 0.5 | 0.5 | 0.0 | 0.5 | 0.5 | 0.0 |
| **Opt. Comb.** |  | 0.6 | 0.3 | 0.1 | 0.2 | 0.8 | 0.0 |

10